\documentclass[12pt]{article}

\usepackage{jheppub}
\usepackage{amsbsy,amsfonts}
\usepackage{graphicx,epsfig}

\newcommand{\p}{\partial}
\newcommand{\del}{\nabla}
\newcommand{\dta}{\delta}
\newcommand{\tr}{\text{tr}}

\title{Chiral vortical effect from the compactified D4-branes with smeared D0-brane charge}
\author[a]{Chao Wu,\footnote{Stay at and become a member of the Wigner Research Center for Physics, Hungarian Academy of Sciences since November 5th 2016.}}
\author[a]{Yidian Chen,}
\author[a,b,c]{Mei Huang}
\affiliation[a]{ Institute of High Energy Physics, Chinese Academy of Sciences, Beijing 100049, P.R. China}
\affiliation[b]{ University of Chinese Academy of Sciences, Beijing 100049, P.R. China}
\affiliation[c]{Theoretical Physics Center for Science Facilities, Chinese Academy of Sciences, Beijing 100049, P.R. China}

\emailAdd{wuchao@ihep.ac.cn}
\emailAdd{chenyd@ihep.ac.cn}
\emailAdd{huangm@ihep.ac.cn}

\abstract{By using the boundary derivative expansion formalism of fluid/gravity correspondence, we study the chiral vortical effect from the compactified D4-branes with smeared D0-brane charge. This background corresponds to a strongly coupled, nonconformal relativistic fluid with a conserved vector current. The presence of the chiral vortical effect is induced by the addition of a Chern-Simons term in the bulk action. Except that the non-dissipative anomalous viscous coefficient and the sound speed rely only on the chemical potential, most of the other thermal and hydrodynamical quantities of the first order depend both on the temperature and the chemical potential. According to our result, the way that the chiral vortical effect coefficient depends on the chemical potential seems irrelevant with whether the relativistic fluid is conformal or not. Stability analysis shows that this anomalous relativistic fluid is stable and the doping of the smeared D0-brane charge will slow down the sound speed.}


\arxivnumber{1608.04922}

\begin{document}

\maketitle

\section{Introduction}

The hot and dense QCD matter has attracted lots of attention recently, because it sheds lights on the mysterious aspects of strongly coupled systems such as the non-trivial topological vacuum, quark confinement and so on. One way to investigate the properties of QCD matter is to measure its responses with respect to external fields such as the electromagnetic fields and the fluid vorticity. Up to the order of linear response, one has
\begin{align}\label{response coefficients}
  J_{(1)}^\mu &=  \sigma_E E^\mu+\sigma_B B^\mu +  \sigma_V l^\mu, \\
  J_{(1)}^{5\mu} &= \sigma_E^5 E^\mu + \sigma^5_B B^\mu + \sigma^5_V  l^\mu.
\end{align}
where $E^\mu=F^{\mu\nu}u_\nu$ and $B^\mu=\frac12\epsilon^{\mu\nu\rho\sigma}u_\nu F_{\rho\sigma}$ are separately the background electric and magnetic field in the rest frame of fluid. $l^\mu=-\epsilon^{\mu\nu\rho\sigma}u_\nu\p_\rho u_\sigma$ is the 4-vorticity vector of the fluid\footnote{The anti-symmetric symbol $\epsilon_{\mu\nu\rho\sigma}$ is defined as $\epsilon_{0123}=-\epsilon^{0123}=1$.}. The subscript ``$(1)$" of $J_{(1)}^\mu$ and $J_{(1)}^{5\mu}$ means that as the responses of the system to $E^\mu,B^\mu$ and $l^\mu$, they are the first order contributions to the total currents from the hydrodynamical point of view.

The response coefficients in eq. (\ref{response coefficients}) contains quite a lot of important and interesting messages of the QCD matter generated in the high energy nucleon collisions (see for example refs. \cite{Huang1509,Kharzeev1511}). They are the electric conductivity (of the Ohm Law) relates with $\sigma_E$, the Chiral Electric Separation Effect coefficient relates with $\sigma_E^5$ \cite{Huang1303,Jiang1409}; the Chiral Magnetic Effect (CME) \cite{Kharzeev0711,Fukushima0808,Kharzeev0907} relates with $\sigma_B$, the Chiral Separation Effect (CSE) with $\sigma^5_B$ \cite{Son0405,Metlitski0505}; the Chiral Vortical Effect (CVE) \cite{Kharzeev0406,Kharzeev0706,Kharzeev1010} with $\sigma_V$ and the Chiral Vortical Separation Effect (CVSE) with $\sigma^5_V$. All these effects are about a net (axial) vector current is induced when the QCD matter (usually in the condition of both chemical and axial chemical are nonzero) is exposed to the external electromagnetic field or the vorticity of the fluid.

The proposal of CVE can be dated back to ref. \cite{Kharzeev0406} where Kharzeev studies the possibilities of the existence for the non-trivial topological domains (the P-odd bubbles) excited in high energy nucleon collisions. The signature for the P-odd bubbles is the asymmetry of the charged pions in the final state. This asymmetry has a polar axis which can be chosen along the angular momentum vector of the system, which is inherited from the initial colliding nucleons. The non-central collision of the initial hadrons causes a net angular momentum transverse to the reaction plane though the larger part of it is carried away by the spectator nucleons. Then Kharzeev and Zhitnitsky \cite{Kharzeev0706} show that both the angular momentum and the magnetic field can induce the charge separation effect when the excited P-odd vacuum domain is at present.

The CVE can be intuitively explained like this. If a hot and dense QCD matter with both $\mu\neq0$ and $\mu_5\neq0$ has an global rotation $\vec \omega$, an effective angular momentum $\vec L$ associated with $\vec\omega$ will be present. Then the chiral quarks with spin $\vec S$ will couple with $\vec L$ in the form of $-\vec S\cdot\vec L$ which is like the spin-orbit coupling in a quantum mechanical problem in atoms. Chiral quarks are allowed to move only along the direction of $\vec L$ because their momentum $\vec p$ should always point in the same (opposite) direction of $\vec S$ for right (left) handed ones. So an effective vector current will be induced in case that there is more positive charged particles inside this QCD matter. In general, the CVE coefficient will have two parts relate separately with the chemical potential and the temperature. Present researches on the properties of CVE coefficient discover that its temperature dependent part will not get corrected when the plasma couples with scalar fields \cite{Golkar1207} but will get corrected when couples with vector fields \cite{Hou1210}.

Though the early stage concepts about CVE was considered in phenomenological models, the first CVE coefficient was found in holographic models \cite{Erdmenger0809,Banerjee0809,Torabian0903} with the CVE term coming from the Chern-Simons (CS) term in the bulk action. All these 3 works were calculated in the frame work of boundary derivative expansion (BDE) formalism of fluid/gravity correspondence \cite{Bhattacharyya0712,Wu1508,Wu1604}, in which refs. \cite{Erdmenger0809,Banerjee0809} work in the charged $AdS_5$ black hole and ref. \cite{Torabian0903} in the STU black hole \cite{Behrndt9810}. Then Son and Sur\'owka \cite{Son0906} prove that this newly found vorticity term is the requirement of quantum anomaly manifested in macroscopic scale. Son et al.'s work actually open's the field of the anomalous or the parity violating hydrodynamics (see for example refs. \cite{Newman0511,Neiman1011,Bhattacharya1105,Jensen1112,Jensen1207}).

The early stage studies of the holographical CVE coefficient \cite{Erdmenger0809,Banerjee0809,Torabian0903} only focus on the CVE coefficient itself. This condition has changed after Son et al.'s work \cite{Son0906} on the anomalous hydrodynamics where the author introduced also the background gauge fields in the classical background of bulk spacetime. The more recent studies relate with the CVE \cite{Amado1102,Landsteiner1107,Landsteiner1312,Megias1304,Kalaydzhyan1102,Gahramanov1203} turn to be in a more comprehensive way. Among these, refs. \cite{Amado1102,Landsteiner1107,Landsteiner1312,Megias1304} use the R-charged $AdS_5$ black hole, but with manual modifications. To be specific, \cite{Amado1102} proposes the Kubo formulae for the CVE coefficient and the conductivities of the background gauge fields and apply them on the R-charged AdS black hole. Then ref. \cite{Landsteiner1107} use these Kubo formulae in some modified R-charged AdS black hole: it contains a gauge-gravity mixed anomaly sector in the bulk action which is responsible for the temperature dependent part of the CVE coefficient.\footnote{There is an interesting research work \cite{Kalaydzhyan1403} which gives a contradictive proposal: The temperature dependent part of CVE coefficient depends on the number of chiral degrees of freedom in the plasma, not the gravitational anomaly.} Ref. \cite{Landsteiner1312} adds an axial vector sector to modify the R-charged AdS black hole background to get the CSE coefficient. The above 3 papers are using the Green-Kubo formalism technically, while the following 3 use the boundary derivative expansion formalism. Ref. \cite{Megias1304} calculates the second order transport coefficients for the anomalous fluid in the model of \cite{Landsteiner1107}. Both refs. \cite{Kalaydzhyan1102,Gahramanov1203} use the STU black hole, but the author prescind from the original physical meanings of the 3 vector charges and give them new interpretations. They study $\sigma_B,\sigma_B^5,\sigma_V$ and $\sigma_V^5$ in \cite{Kalaydzhyan1102} and $\sigma_B$ with $v_2$ correction in the case of anisotropy hydrodynamics in \cite{Gahramanov1203}.

The holographic studies that we have mentioned on CVE coefficient \cite{Erdmenger0809,Banerjee0809,Torabian0903,Amado1102,Landsteiner1107,Landsteiner1312,Megias1304,Kalaydzhyan1102} use either the R-charged AdS black hole or the STU black hole background, the boundary fluid of which are all conformal. In this paper, based on the nonconformal generalization of the fluid/gravity correspondence constructed via the compactified D4-branes \cite{Wu1508,Wu1604}, we would like to begin the venture for the nonconformal anomalous hydrodynamics. The study of nonconformal fluid is important in that it will help us to recover new information of hydrodynamics. For example, \cite{Kleinert1610} discovers that some relations between the second order transport coefficients of nonconformal fluid proposed in \cite{Kanitscheider0901} are wrong with the help of the results of \cite{Wu1604}. In this paper, we will only study the CVE coefficient and other thermal and hydrodynamical quantities up to the first order as the first step towards the nonconformal anomalous hydrodynamics. In order to introduce the background vector field into the compactified D4-brane, we will use the background of compactified black D4-branes with D0-branes smeared uniformly on its volume. If one makes a double Wick rotation on both the time direction and one of the direction of D4-branes' world volume, then this background will become the D0-D4 Sakai-Sugimoto model \cite{Wu1304,Seki1304}. This model is constructed by adding the smeared D0-brane charge into the Sakai-Sugimoto (SS) model \cite{Sakai0412,Sakai0507} background and it has been used to explore many aspects of QCD holographically \cite{Wu1304,Seki1304,Cai1410,Hanada1505,Bigazzi1506,Li1506,Li1602}.

The organization of this paper is as follows: In section \ref{sec: setup} we will give a clear explanation on the background of compactified D4-brane with smeared D0-brane charge and its dimensional reduction to 5D form. Then in section \ref{sec: perturbation}, we will solve all the perturbations. We will calculate all the thermodynamic and hydrodynamic quantities in section \ref{sec: result}. We will end this paper in section \ref{sec: final} by discussing some problem and some working directions for the future.

\section{The setup}\label{sec: setup}

In this section, we will introduce the background of compactified D4-brane with smeared D0-brane charge and reduce it into 5 dimensional form. This technic has been used firstly in ref. \cite{Benincasa0605} where Benincasa et al. derive the sound speed and $\zeta/\eta$ for the compactified black D4-brane background, and later in refs. \cite{Wu1508,Wu1604} where the authors try to offer a nonconformal counterpart to Bhattacharyya et al.'s $AdS_5$ construction of fluid/gravity correspondence \cite{Bhattacharyya0712}.

The 10D effective action for the type IIA superstring theory with both D0 and D4-brane at present reads as:
\begin{align}\label{D0-D4 action}
  S = \frac1{2\kappa_{10}^2} \int d^{10}x\sqrt{-G}\left[ R^{(10)} - \frac12(^{10}\del\phi)^2 - \frac{g_s^2}{2\cdot4!} e^{\frac{\phi}2} F_4^2 - \frac{g_s^2}{2\cdot2!} e^{\frac32\phi} F_2^2 \right],
\end{align}
The background for $N_4$ compactified black D4-branes with $N_0$ smeared D0-branes is
\begin{align} \label{D0-D4 bakgrd fields}
  ds^2 &=  - H_0^{-\frac78}H_4^{-\frac38}f(r)dt^2 + H_0^\frac18H_4^{-\frac38}(d\vec x^2 + dy^2) + H_0^\frac18 H_4^{\frac58}\left( \frac{dr^2}{f(r)} + r^2 d\Omega_4^2\right), \cr
  e^\phi &= e^{\Phi-\Phi_0} = H_0^{\frac34}H_4^{-\frac14},~~~F_4 = g_s^{-1}Q_4\epsilon_4,~~~F_2 = dA_1 = \frac{g_s^{-1}Q_0}{r^4H_0^2}dr\wedge dt,
\end{align}
with
\begin{align}
  f(r) = 1-\frac{r_H^3}{r^3},~~~H_0 = 1 + \frac{r_{Q0}^3}{r^3},~~~H_4 = 1 + \frac{r_{Q4}^3}{r^3},~~~A_1 = g_s^{-1}\sqrt{1+\frac{r_H^3}{r_{Q0}^3}}(H_0^{-1}-1)dt.
\end{align}
Here $\Phi$ is the dilaton with $\Phi_0$ its vacuum value and $g_s$ is the string coupling defined as $g_s=e^{\Phi_0}$. $A_1$ is the Ramond-Ramond (RR) field coupled with D0-branes, with $F_2$ its field strength. $F_4$ is the RR field strength magnetically coupled with the D4-branes. The $r_H$, $r_{Q0},r_{Q4}$, $Q_0$ and $Q_4$ are the parameters of this background. The $Q_0$ and $Q_4$ can be given as \footnote{Here the $\Omega_4$ is the volume of unit 4-sphere and $V_4$ is the spatial volume of the D4-brane.}
\begin{align}\label{expressions of Q0 and Q4}
  Q_0 = \frac{(2\pi l_s)^7g_sN_0}{V_4\Omega_4},~~~~Q_4 = \frac{(2\pi l_s)^3g_sN_4}{\Omega_4},
\end{align}
by using the normalization conditions of the RR fields. They relate with $r_H$ and $r_{Q0},r_{Q4}$ via EOM as
\begin{align}\label{Q_0 and Q_4}
  Q_0^2 = 9r_{Q0}^3( r_{Q0}^3 + r_H^3 ),~~~~Q_4^2 = 9r_{Q4}^3( r_{Q4}^3 + r_H^3 ).
\end{align}
Under the near horizon limit, $H_4\to L^3/r^3$ with $L^3=\pi g_sN_4l_s^3$ thus the background metric and dilaton becomes
\begin{align}
  ds^2 &= -\left(\frac rL\right)^\frac98 \left( H_0^{-\frac78} fdt^2  + H_0^\frac18 (d\vec x^2+dy^2) \right) + \left(\frac Lr\right)^\frac{15}8 H_0^\frac18 \left( \frac{dr^2}{f} + r^2d\Omega_4^2 \right), \label{10D Einstein metric in near horizon limit} \\
  e^\phi &= \left(\frac rL\right)^\frac34 H_0^\frac34.
\end{align}
The D4-branes are lying in directions of $\{x^i,y\}$ with $y$ compact hence the name compactified D4-brane. So the above metric is the near-extremal, compactified D4-branes with smeared D0-branes with the topology is $\mathbf D_2\times\mathbf R^3\times\mathbf S^1\times\mathbf S^4$, where $\mathbf D_2$ is the 2D disk of $\{r,t\}$ surface.

If one makes Wick rotations on both $dt$ and $dy$ at the same time: $dt^2\to-d\tau^2,dy^2\to-(dx^0)^2$ with $x^0$ noncompact, then the 10D Einstein metric in the near horizon limit becomes the D0-D4 Sakai-Sugimoto model metric:
\begin{align}
  ds^2 &= \left(\frac rL\right)^\frac98 \left( H_0^\frac18 \eta_{\mu\nu}dx^\mu dx^\nu + H_0^{-\frac78} fd\tau^2 \right) + \left(\frac Lr\right)^\frac{15}8 H_0^\frac18 \left( \frac{dr^2}{f} + r^2d\Omega_4^2 \right), \cr
   f(r) &= 1-\frac{r_{KK}^3}{r^3}.
\end{align}
This metric is a bubble configuration with topology of $\mathbf R_4\times \mathbf D_2\times \mathbf S^4$ with $\mathbf D_2$ the 2D disk of $\{r,\tau\}$. In the bubble background the D0-brane's RR field is $A_\tau$, which is a spatial component. Since the spatial component of a vector is a pseudoscalar. So it can be interpreted \cite{Sakai0412} that its integration along $\tau$ corresponds to $\theta$: $\theta=\int d\tau A_\tau=\iint d\tau dr F_{r\tau}$. The reason that $A_\tau$ can be interpreted as the theta angle lies on this: In the SS model, the field theory is not on the boundary, it is on the world volume of D4-brane. The action of the effective field theory on the world-volume of D4-brane has a term as
\begin{align}
  S_{CS}^{(D4)} = \frac12 \mu_4 (2\pi\alpha')^2 \int d\tau A_\tau \tr{\mathcal F_2 \wedge \mathcal F_2}.
\end{align}
Since one has $\theta=\int d\tau A_\tau$ thus $S_{CS}^{(D4)}\sim\theta\int\tr{\mathcal F_2 \wedge \mathcal F_2}$ which is the $\theta$ term in field theory. Here $\mathcal F_2$ is the gauge field strength on the world-volume of D4-brane (not D0-brane's RR field strength $F_2$). But the 10D metric we use is of black brane type, not of bubble type. The D0-brane's RR field $A_t$ is in the real time direction $dt$ now, it is just a scalar so we can not relate it with $\theta$ any longer. Thus $A_1$ in our paper is a vector and its nonzero component in the background $A_t$ can be related with chemical potential of the corresponding fluid.

We use the following ansatz to reduce the metric into 5D form as in refs. \cite{Benincasa0605,Wu1508} ($L=1$ from now on):
\begin{align}\label{dim reduction ansatz}
  ds^2 = e^{-\frac{10}3A}g_{MN}dx^Mdx^N + e^{2A+8B}dy^2 + e^{2A-2B}d\Omega_4^2.
\end{align}
The reduced 5D background will be
\begin{align}\label{5d reduced metrc}
  ds^2 &= - r^\frac53H_0^{-\frac23}(r)f(r)dt^2 + r^\frac53H_0^{\frac13}(r)d\vec x^2 + \frac{H_0^\frac13(r)}{r^\frac43f(r)}dr^2.\cr
  e^\phi &= r^\frac34 H_0^\frac34, ~~e^A = r^\frac{13}{80}H_0^\frac1{16},~~e^B = r^\frac1{10},~~A_1 = g_s^{-1}\sqrt{1+\frac{r_H^3}{r_{Q0}^3}}(H_0^{-1}-1)dt.
\end{align}
As one can check that the scalar curvature and the square of Riemann tensor behave like $1/r^{2/3}$ and $1/r^{4/3}$, respectively near the boundary. Thus the metric of the 5D reduce background (\ref{5d reduced metrc}) is asymptotically flat which is the same as in ref. \cite{Wu1508}. The Hawking temperature is
\begin{align}\label{Hawking temperature}
  T = \frac{3r_H^\frac12}{4\pi H_0^\frac12(r_H)} = \frac{3r_H^2}{4\pi(r_H^3+r_{Q0}^3)^\frac12}.
\end{align}
The expression for $Q_4$ in eq. (\ref{expressions of Q0 and Q4}) gives $Q_4=3\pi g_sN_4l_s^3=3L^3$. Since we have set $L=1$, thus $Q_4=3$. Here we would like to define a parameter relates with $Q_0$ as
\begin{align}
  n_0^2 \equiv \frac{Q_0^2}{Q_4^2} = r_{Q0}^3(r_{Q0}^3+r_H^3).
\end{align}
Given that $Q_4=3$, one has $Q_0 = 3n_0$, which will be used later.
%

The 5D reduced bulk action can be got through the following procedures:
\begin{align}
  \sqrt{-G} &= \sqrt{-g}\sqrt\gamma e^{-\frac{10}3A},\\
  \sqrt{-G}R^{(10)} &= \sqrt{-g}\sqrt\gamma\left(R + \frac{10}3\del^2A - \frac{40}3(\p A)^2 - 20(\p B)^2 + 12e^{-\frac{16}3A + 2B}\right), \\
  \sqrt{-G}(^{10}\del\phi)^2 &= \sqrt{-g}\sqrt\gamma e^{-\frac{10}3A}G^{\hat M\hat N}\p_{\hat M}\phi\p_{\hat N}\phi=\sqrt{-g}\sqrt\gamma e^{-\frac{10}3A} (e^{\frac{10}3A}g^{MN})\p_M\phi\p_N\phi \cr
  & = \sqrt{-g}\sqrt\gamma(\p\phi)^2, \\
  \sqrt{-G}\frac{g_s^2}{2\cdot4!}e^{\frac{\phi}2}F_4^2 &= \sqrt{-g}\sqrt\gamma\frac{Q_4^2}2e^{\frac\phi2-\frac{34}3A+8B}, \\
  \sqrt{-G}\frac{g_s^2}{2\cdot2!}e^{\frac{3\phi}2}F_2^2 &= \sqrt{-g}\sqrt\gamma e^{-\frac{10}3A} \frac{g_s^2}{2\cdot2!} e^{\frac{3\phi}2} (e^{\frac{10}3A}g^{MP}) (e^{\frac{10}3A}g^{NQ}) F_{MN}F_{PQ} \cr
  & = \sqrt{-g}\sqrt\gamma \frac{g_s^2}{2\cdot2!} e^{\frac32\phi+\frac{10}3A}F_{MN}^2,
\end{align}
where the indices with a ``hat" like $\hat M,\hat N$ are 10 dimensional ones and those without a ``hat" like $M,N$ are 5 dimensional ones. Here $g=\det{g_{MN}}$ and $\gamma$ is the determinant of the matric on $S^4$. We write an explicit superscript ``$(10)$" on some quantities to indicate they are 10D quantities. The details of the reduction of $R^{(10)}$ to its 5D form $R$ can be found in the appendix of ref. \cite{Wu1508}. Thus the reduced 5D bulk action is
\begin{align}
  S&=\frac1{2\kappa_5^2}\int d^5x\sqrt{-g}\bigg[ R - \frac12(\p\phi)^2 - \frac{40}3(\p A)^2 - 20(\p B)^2 - V(\phi,A,B) - \frac{g_s^2}4e^{\frac32\phi + \frac{10}3A}F_{MN}^2 \bigg], \cr
  V&(\phi,A,B)=\frac{Q_4^2}2e^{\frac\phi2-\frac{34}3A+8B}-12e^{-\frac{16}3A+2B},
\end{align}
where $\frac1{2\kappa_5^2}=\frac{\Omega_4\beta_y}{2\kappa_{10}^2}$ is the 5 dimensional surface gravity and $\beta_y=\int dy$ is the circumference of $S^1$. Compared with the case of ref. \cite{Wu1508}, here the action receives the contribution from the D0-branes' RR field, i.e. $A_M$. Thus this system is 5D Einstein gravity coupled with 3 scalars and a vector field. The dual field theory will have a chemical potential as we will see in the final results.

According to refs. \cite{Wu1508,Wu1604}, the full action of the reduced 5D system is
\begin{align}\label{5D action: total}
  S=S_{bulk}-\frac1{\kappa_5^2}\int d^4x\sqrt{-h}K+\frac1{\kappa_5^2}\int d^4x\sqrt{-h}\frac52e^{-\frac53A-\frac1{12}\phi},
\end{align}
where the second term in the r.h.s. of the above equation is the Gibbons-Hawking term and the third term is the counter term. The 5 dimensional bulk action is
\begin{align}\label{5D action: bulk}
  S_{bulk} &= \frac1{2\kappa_5^2}\int d^5x\bigg\{ \sqrt{-g}\bigg[ R - \frac12(\p\phi)^2 - \frac{40}3(\p A)^2 - 20(\p B)^2 - V(\phi,A,B) \cr
                    & - \frac{g_s^2}4e^{\frac32\phi + \frac{10}3A}F_{MN}^2 \bigg] + \frac13g_s^3\kappa_{CS}\epsilon^{MNPQR}A_M F_{NP}F_{QR}\bigg\}, \cr
  V(\phi,A,&B) = \frac92e^{\frac\phi2-\frac{34}3A+8B}-12e^{-\frac{16}3A+2B}.
\end{align}
Here we define the Levi-Civita symbol as it is in flat spacetime with the metric $\eta_{MN}=\text{diag}\{-1,1,1,1,1\}$ with the convention $\epsilon^{01234}=-\epsilon_{01234}=-1$. $\kappa_{CS}$ is the coupling of the CS term. Note that in the above bulk action, we add manually a CS term for the D0-branes RR field which corresponds to the vorticity term in the dual relativistic fluid. This CS term does not have a 10D origins. It is added just ``by hand". Here we take a similar viewpoint as \cite{Landsteiner1107,Megias1304,Landsteiner1312,Kalaydzhyan1102} that we will not be very strict on the 10D string theory origin of the 5D reduced theory with the full action is (\ref{5D action: total}). Generally speaking, the technic of BDE formalism does not relate with the 10D string theory directly but rely more on calculating the Brown-York tensor for the 5D reduced background. So one may take the same standpoint as \cite{Kalaydzhyan1102} that to view the 5D reduced system as a bottom-up holographic model which does not have direct relations with its 10D origins. In practical application, the manually added CS term for the D0-brane RR field does not couple with the scalar fields or the metric tensor. So it will not change any properties of the scalar part or the tensor part of the 5D theory. It may only modify, if it will, the topological property of the vector field since it is a topological term. But we do not use the topological property of the vector field in the calculations. So pragmatically, we think this is enough to justify the manually added CS term in the 5D reduced theory.

The EOM can be derived out from eq. (\ref{5D action: bulk}) as
\begin{align}
  E_{MN}&-T_{MN}=0,\label{EOM: Ein} \\
  \del^2\phi& - \frac94e^{\frac\phi2-\frac{34}3A+8B} - \frac38g_s^2e^{\frac32\phi+\frac{10}3A}F_{MN}^2=0, \label{EOM: phi}\\
  \del^2A& + \frac{153}{80}e^{\frac\phi2-\frac{34}3A+8B} - \frac{12}5e^{-\frac{16}3A+2B}- \frac1{32}g_s^2e^{\frac32\phi+\frac{10}3A}F_{MN}^2 =0,\label{EOM: A} \\
  \del^2B& - \frac{9}{10}e^{\frac\phi2-\frac{34}3A+8B} + \frac35e^{-\frac{16}3A+2B}=0, \label{EOM: B}\\
  \p_N( g_s^2&\sqrt{-g}e^{\frac32\phi+\frac{10}3A}F^{MN}) - g_s^3\kappa_{CS}\epsilon^{MNPQR} F_{NP}F_{QR}=0. \label{EOM: RR field}
\end{align}
In the above equation, $E_{MN}$ is the Einstein tensor and it is defined as
\begin{align}
  E_{MN}\equiv R_{MN}-\frac12g_{MN}R;
\end{align}
$T_{MN}$ is the energy-momentum tensor in the 5D bulk defined as
\begin{align}
  T_{MN}&\equiv \frac12\left(\p_M\phi\p_N\phi-\frac12g_{MN}(\p\phi)^2\right)+\frac{40}3\left(\p_M A\p_N A-\frac12g_{MN}(\p A)^2\right)\cr
  &+20\left(\p_M B\p_N B-\frac12g_{MN}(\p B)^2\right) +\frac{g_s^2}2e^{\frac32\phi+\frac{10}3A}\left(F_{MP}F_N^{~P}-\frac14g_{MN}F_2^2\right) \cr
  & - \frac12g_{MN}V.
\end{align}

As in ref. \cite{Bhattacharyya0712}, we boost the reduced 5D background by $dv\to-u_\mu dx^\mu,~dx^i\to P^i_{~\mu}dx^\mu$ and (\ref{5d reduced metrc}) becomes
\begin{align}\label{5D boosted background fields}
  ds^2 &= - r^\frac53H_0^{-\frac23}f(r)u_\mu u_\nu dx^\mu dx^\nu + r^\frac53H_0^{\frac13}P_{\mu\nu}dx^\mu dx^\nu - 2r^\frac16 H_0^{-\frac16}u_\mu dx^\mu dr, \cr
  e^\phi &= r^\frac34 H_0^\frac34, ~~~~e^A = r^\frac{13}{80}H_0^\frac1{16},~~~~e^B = r^\frac1{10}, \cr
  A_1 &= g_s^{-1}\sqrt{1+\frac{r_H^3}{r_{Q0}^3}}(1-H_0^{-1})u_\mu dx^\mu,~~~~F_2 = dA_1 = \frac{g_s^{-1}Q_0}{r^4H_0^2}u_\mu dx^\mu,
\end{align}
where
\begin{align}
  f(r) = 1-\frac{r_H^3}{r^3},~~~~H_0(r) = 1 + \frac{r_{Q0}^3}{r^3},~~~~u^\mu = \frac{(1,\beta_i)}{\sqrt{1-\beta_i^2}}.
\end{align}
Now we let the parameters in the above equation i.e. $r_H$, $r_{Q0}$ and $\beta_i$ to be slowly $x^\mu$-dependent, which means $\left|\frac{\p_\mu\#}T\right|\ll1,~\#=\{r_H,r_{Q0},\beta_i\}$. The physical meaning of $r_H$ and $\beta_i$ to be boundary coordinate dependent has been explained clearly in refs. \cite{Bhattacharyya0712,Wu1508}. We would like to give an explanation to $x^\mu$ dependence of $r_{Q0}$. From eq. (\ref{Q_0 and Q_4}) one can see that $r_{Q0}$ relates with $n_0$ and $r_H$, thus $r_{Q0}$ is $x^\mu$ dependent is equal to $n_0$ is $x^\mu$ dependent. We know that $n_0$ is the relative density of D0-branes, thus $n_0$ is $x^\mu$ dependent means the D0 charge is no longer uniform---it has fluctuations which should be in the long wavelength limit due to the condition: $\left|\frac{\p_\mu r_{Q0}}T\right|\ll1$. The gradients of D0-brane density and the CS term of $A_M$ will behave like two sources that separately contributes to the first derivative order of the conserved vector current $J^\mu$, as will be seen in the final results of this paper. Though $n_0$ has a clearer physical significance than $r_{Q0}$, we will still use $r_{Q0}$ in the calculation of the following sections since it will make the formulations look more neatly.

\section{Solving the perturbations}\label{sec: perturbation}

We expand the $x^\mu$ dependent metric in eq. (\ref{5D boosted background fields}) to first order as
\begin{align}\label{5D expanded metric}
  ds^2 =& -r^\frac53H_0^{-\frac23}\left(f - \frac{3r_H^2\dta r_H}{r^3} - \frac{2fr_{Q0}^2\dta r_{Q0}}{r^3H_0}\right)dv^2 + 2r^\frac53H_0^{-\frac23}(f-H_0)\dta\beta_idx^idv \cr
  & + 2r^\frac16H_0^{-\frac16}\left(1 - \frac{r_{Q0}^2\dta r_{Q0}}{2r^3H_0}\right)dvdr + r^\frac53H_0^\frac13\left(1 + \frac{r_{Q0}^2\dta r_{Q0}}{r^3H_0}\right)d\vec x^2 \cr
  & - 2 r^\frac16H_0^{-\frac16}\dta\beta_idx^idr.
\end{align}
where $\dta\#=x^\mu\p_\mu\#$, with $\#=\{r_H,r_{Q0},\beta_i\}$. We set the perturbations of the metric as
\begin{align}
  ds^2 =& -r^\frac53H_0^{-\frac23}k(x,r)u_\mu u_\nu dx^\mu dx^\nu + 2r^\frac53H_0^{-\frac23}P_\mu^\rho w_\rho(x,r)u_\nu dx^\mu dx^\nu \cr
  & + r^\frac53H_0^\frac13(\alpha_{\mu\nu}(x,r)+h(x,r)P_{\mu\nu})dx^\mu dx^\nu - 2r^\frac16H_0^{-\frac16}j(x,r)u_\mu dx^\mu dr.
\end{align}
To first order, it becomes
\begin{align}\label{1st order pert ansatz}
  ds^2 =& - r^\frac53H_0^{-\frac23}k^{(1)}(r)dv^2 - 2r^\frac53H_0^{-\frac23} w_i^{(1)}(r)dx^i dv + 2 r^\frac16H_0^{-\frac16} j^{(1)}(r)dvdr \cr
  & + r^\frac53H_0^\frac13(\alpha^{(1)}_{ij}(r) + h^{(1)}(r)\dta_{ij})dx^idx^j.
\end{align}

The vector field together with the preset perturbations reads as
\begin{align}
  A_1 = g_s^{-1}\sqrt{1+\frac{r_H^3(x)}{r_{Q0}^3(x)}}(1-H_0^{-1}(x,r))u_\mu(x) dx^\mu + P_\mu^\nu a_\nu(x,r)dx^\mu + c(x,r)u_\mu(x) dx^\mu.
\end{align}
To first order, the above becomes
\begin{align}\label{1st order axial vec with pert}
  A_1 =& - g_s^{-1}\left[\left(\frac{n_0}{r_{Q0}^3} + \frac{3r_H^2\dta r_H}{2n_0} - \frac{3r_H^3\dta r_{Q0}}{2n_0r_{Q0}}\right)(1 - H_0^{-1}) + \frac{3n_0\dta r_{Q0}}{r^3H_0^2r_{Q0}}\right]dv \cr
   & + \frac{g_s^{-1}n_0}{r_{Q0}^3}(1-H_0^{-1})\dta\beta_idx^i - c^{(1)}(r)dv + a^{(1)}_i(r)dx^i.
\end{align}

Since from (\ref{5d reduced metrc}) one can see that the scalar field $\phi$ and $A$ (should not be confused with D0-brane's RR field $A_M$) contains $H_0$ thus will also be $x^\mu$ dependent after the promotion of $r_{Q0}$ to be boundary coordinate dependent. So after the derivative expansion, $\phi$ and $A$ become
\begin{align}\label{phi and A with 1st pert}
  \phi = \ln\left[  r^\frac34 H_0^\frac34 \left( 1+\frac{9r_{Q0}^2 \dta r_{Q0}}{4H_0r^3} \right) \right], ~~~~A = \ln\left[ r^\frac{13}{80} H_0^\frac1{16} \left( 1+\frac{3r_{Q0}^2 \dta r_{Q0}}{16H_0r^3} \right) \right].
\end{align}
Note that here we do not turn on any perturbations for the scalar fields. This makes the EOMs of the scalar fields contains only the scalar part perturbations of metric tensor and vector field. One can of course turn on perturbations for $\phi$, $A$ and $B$, for which we may leave as future studies. The $x^\mu$ dependence of scalar fields will also modify the Brown-York tensor, we will see in section 4.

In the rest of this section, we will put the metric (\ref{5D expanded metric}) with its perturbations (\ref{1st order pert ansatz}) together with the vector field (\ref{1st order axial vec with pert}) and scalar fields (\ref{phi and A with 1st pert}) upto first order into the EOMs of the 5D system to solve all the perturbations out in the asymptotic regime, i.e. near the boundary. We set $g_s=1$ from now on and will omit the superscript ``$(1)$" for all the first order perturbation ansatz.

\subsection{The tensor part}

The EOM of the tensor part is
\begin{align}
  E_{ij} - \frac13\delta_{ij}\delta^{kl}E_{kl} = T_{ij}-\frac13\delta_{ij}\delta^{kl}T_{kl}.
\end{align}
By substituting the expanded metric and the metric perturbations one has the differential equation of $F(r)$
\begin{align}\label{diff eq of F(r)}
  \p_r(r^4f\p_rF) = - \frac{5r^3 + 2r_{Q0}^3}{r^\frac32H_0^\frac12},
\end{align}
where $F(r)$ satisfies $\alpha_{ij}=F(r)\sigma_{ij}$, with $\sigma_{ij}=\p_{(i}\beta_{j)}-\frac13\dta_{ij}\p\beta$ the spatial part of the shear viscous tensor.

The solution of eq. (\ref{diff eq of F(r)}) can be written formally as
\begin{align}
  F(r) = \int_\infty^r \frac1{x^4f(x)}dx \int_{r_H}^x \left( -\frac{5y^3 + 2r_{Q0}^3}{y^\frac32H_0^\frac12(y)} \right)dy.
\end{align}
We only need the asymptotic behavior of the result for the above integral. So we expand the above in terms of $1/r$ and get
\begin{align}
  F(r) = \frac4{r^\frac12} - \frac{2r_H\sqrt{r_H^3+r_{Q0}^3}}{3r^3}.
\end{align}
More details about this integral can be found in refs. \cite{Bhattacharyya0712,Wu1604}.

\subsection{The vector part}

The vector part is quite different from the case in refs. \cite{Wu1508,Wu1604}, here it has two sectors of perturbations: $w_i$ from the metric and $a_i$ from the vector field. The differential equations for them are coupled with each other, but can be decoupled and solved independently.

The constraint equation of vector part from the Einstein equation is
\begin{align}
  g^{r0}(E_{0i}-T_{0i})+g^{rr}(E_{ri}-T_{ri})=0,
\end{align}
which gives
\begin{align}
  \frac{r_H^2}{r_H^3+r_{Q0}^3}\p_ir_H = -2\p_0\beta_i.
\end{align}
One can check that if $r_{Q0}=0$, we will have $\p_ir_H= -2r_H\p_0\beta_i$ which is the first order vector constraint in the compactified D4-brane case \cite{Wu1508}. The dynamical equation from the Einstein equation is
\begin{align}
  E_{ri}-T_{ri}=0,
\end{align}
this gives
\begin{align}\label{Einstein Eq (ri)}
  & 4r^8H_0^2w_i'' + 4r^4H_0(4r^3+7r_{Q0}^3)w_i' + 36r_{Q0}^6w_i -12n_0r^4H_0^2a_i' \cr
  & - 3r^{-\frac12} H_0^\frac12 (r^3 + 4r_{Q0}^3) r_{Q0}^2 \p_ir_{Q0} - 2r^\frac52H_0^\frac32(5r^3+2r_{Q0}^3)\p_0\beta_i = 0.
\end{align}

The $(i)$ component of the Maxwell equation for the vector field gives
\begin{align}\label{Maxwell Eq (i)}
  & \p_r(r^4H_0fa_i') - \bigg( \frac{3n_0}{H_0}w_i' + \frac{9n_0r_{Q0}^3}{r^4H_0^2}w_i \bigg) + \frac{n_0(r^3+4r_{Q0}^3)}{r^\frac92H_0^\frac12}\p_0\beta_i \cr
  & -\frac{3[n_0^2(r^6-9r_{Q0}^3r^3-4r_{Q0}^6) + r_{Q0}^6r^3H_0(r^3+4r_{Q0}^3)]}{ 4n_0r_{Q0}r^\frac{15}2H_0^\frac32 }\p_ir_{Q0} + \frac{24\kappa_{CS}n_0^2}{r^7H_0^3}l_i = 0,
\end{align}
where $l_i=\epsilon_{ijk}\p_j\beta_k$ is the spatial component of $l_\mu$. From eq. (\ref{Einstein Eq (ri)}) and eq. (\ref{Maxwell Eq (i)}) we can eliminate $a_i$ and get the quation for $w_i$ as
\begin{align}
  \p_r[r^4f\p_r(r^4w_i')] &= \frac{3r_{Q0}^2}{8r^\frac{15}2 H_0^\frac52} \Big[5r^9 + (7r_H^3 + 16r_{Q0}^3)r^6 - 4r_{Q0}^3(7r_H^3+4r_{Q0}^3)r^3  \cr
  & - 8r_H^3r_{Q0}^6 \Big] \p_ir_{Q0} + \frac1{4r^\frac92 H_0^\frac32} \Big[ 55r^9 - 25(r_H^3 - 2r_{Q0}^3)r^6  \cr
  & - 4r_{Q0}^3 (5r_H^3 + 2r_{Q0}^3) r^3 - 8r_{Q0}^6 (5r_H^3 + 6r_{Q0}^3) \Big] \p_0\beta_i - \frac{72\kappa_{CS}n_0^3}{r^7H_0^4} l_i.
\end{align}
The solution of the differential equation for $w_i$ is
\begin{align}
  w_i(r) = -\frac2{r^\frac12}\p_0\beta_i + \mathcal O\left(\frac1{r^\frac72}\right) \p_ir_{Q0} + \mathcal O\left(\frac1{r^6}\right)l_i,
\end{align}
where we only record terms with the order less than $\mathcal O\left(\frac1{r^3}\right)$ here, but in the calculation process higher order terms should be taken into consideration so that one can get the correct solution for $a_i$, which is
\begin{align}
  a_i(r) = - \frac{4r_H r_{Q0}^\frac32}{3r^3} \p_0\beta_i - \frac{r_H^4 r_{Q0}^\frac12}{2(r_H^3 + r_{Q0}^3) r^3} \p_ir_{Q0} + \frac{4\kappa_{CS}r_{Q0}^3(3r_H^3+r_{Q0}^3)}{9(r_H^3+r_{Q0}^3)^2r^3}l_i.
\end{align}
Note that $w_i$ still does not have $1/r^3$ order terms as in \cite{Wu1508}, which means it will not contribute to the conserved vector current of the boundary fluid. The vector part perturbation $a_i$ of $A_M$ will contribute to the conserved current since it contains the $1/r^3$ order terms. The interesting point is, though the CS term of $A_M$ is added by hand, it turns out to have a physical contribution as can be seen from the term of $l_i$.

\subsection{The scalar part}

From refs. \cite{Wu1508,Wu1604} we learn that to nonconformal fluid in the prescription of BDE formalism of fluid/gravity correspondence, the scalar part is the most complicate. In the situations considered in this paper, the scalar part is even more complex than that in refs. \cite{Wu1508,Wu1604}. We will separate the constraint and the dynamical equation into two parts to study.

\subsubsection{The constraint equations}

There is one more constraint equations of scalar part than it is in refs. \cite{Wu1508,Wu1604}, which is the $(r)$ component of the Maxwell equations for the vector field.

The first scalar constraint from the Einstein equation is
\begin{align}
  g^{rr}(E_{r0}-T_{r0})+g^{r0}(E_{00}-T_{00})=0,
\end{align}
which gives
\begin{align}
  r_H^2(5r^3+2r_{Q0}^3)\p_0r_H + 3r_{Q0}^2(2r^3-r_H^3)\p_0r_{Q0} + 2r^3(r_H^3+r_{Q0}^3)\p\beta =0.
\end{align}
The $(r)$ component of Maxwell equation gives
\begin{align}
  3r_H^2r_{Q0}\p_0r_H + 3(r_H^3+2r_{Q0}^3)\p_0r_{Q0} + 2r_{Q0}(r_H^3+r_{Q0}^3)\p\beta = 0.
\end{align}
The above two constraint equations contain no scalar perturbations, from which we can solve $\p_0r_H$ and $\p_0r_{Q0}$ out in terms of $\p\beta$:
\begin{align}
  \frac1{r_H}\p_0r_H = -\frac{2(r_H^3+r_{Q0}^3)}{5r_H^3+4r_{Q0}^3}\p\beta, ~~~~\frac1{r_{Q0}}\p_0r_{Q0} = -\frac{4(r_H^3+r_{Q0}^3)}{3(5r_H^3+4r_{Q0}^3)}\p\beta.
\end{align}
One can check that if setting $r_{Q0}=0$, we will have $\frac1{r_H}\p_0r_H = -\frac25\p\beta$, which is the first scalar constraint in ref. \cite{Wu1508}; the other constraint will become a trivial identity. The above two constraint relations will be useful in tackling the following equations in that one can use it to change the sources of both $\p_0r_{Q0}$ and $\p_0r_H$ into the source of $\p\beta$.

The second scalar constraint from Einstein equation is
\begin{align}
  g^{rr}(E_{rr}-T_{rr})+g^{r0}(E_{r0}-T_{r0})=0,
\end{align}
which gives
\begin{align}\label{Diff Eq: 2nd Einstein constraint}
  &r^4H_0(5r^3+2r_{Q0}^3)k' + 3(5r^6+10r_{Q0}^3r^3+2r_{Q0}^6)k - 30r^6H_0^2j + 3r^4H_0^2(5r^3-2r_H^3)h' \cr
  & - 6n_0r^4H_0^2c' + \left[ 2r^\frac52H_0^\frac32(5r^3+2r_{Q0}^3) + \frac{4n_0^2H_0^\frac12(r^3-2r_{Q0}^3)}
  {(5r_H^3+4r_{Q0}^3)r^{\frac12}} \right]\p\beta = 0.
\end{align}
We have changed the source term with $\p_0r_{Q0}$ into $\p\beta$, we will continue to do this for the following dynamical equations without pointing it out again. The above equation contains the first order derivative of perturbations, and will be usefull when we solving them.

\subsubsection{The dynamical equations}

There are two main differences for the dynamical equations of scalar perturbations compared with ref. \cite{Wu1508}. The first one is that we should take
the $(0)$ component of the Maxwell equation into consideration. The second one is, the EOMs of the three scalars $\phi$, $A$ and $B$ produce the same differential equations in \cite{Wu1508} but here their EOMs produce different differential equations. So in general, we should consider 6 dynamical equations. They are the $(rr)$ and $(ii)$ (with $i$ summed) components of Einstein equations, the EOMs for the three scalars and the $(0)$ component of the Maxwell equation.

The $(rr)$ component of Einstein equation is
\begin{align}
  E_{rr}-T_{rr}=0,
\end{align}
after putting into the first order expanded metric together with the perturbations, this gives
\begin{align}
  6r^4H_0h'' + 9r^3h' -2(5r^3+2r_{Q0}^3)j' = 0.
\end{align}
This is the most simple one among those 6 dynamical equations.

The EOM of $\phi$ (\ref{EOM: phi}) gives
\begin{align}\label{Diff Eq: phi}
  & 2r^4H_0(r^3-2r_{Q0}^3)k' + 6(r^6+2r_{Q0}^3r^3-2r_{Q0}^6)k - 2r^4H_0f(r^3-2r_{Q0}^3)j' - 12r^6H_0^2j \cr
  &  + 3r^4 H_0 f (r^3 - 2r_{Q0}^3) h' - 12n_0r^4H_0^2c' + \left[ 2r^\frac52H_0^\frac32(r^3-2r_{Q0}^3) + \frac{24n_0^2 r^{\frac52}H_0^\frac12}{5r_H^3+4r_{Q0}^3} \right] \p\beta = 0,
\end{align}
the EOM of $A$ (\ref{EOM: A}) gives
\begin{align}\label{Diff Eq: A}
  & 2r^4H_0(13r^3-2r_{Q0}^3)k' + 6(13r^6+26r_{Q0}^3r^3-2r_{Q0}^6)k - 2r^4H_0f(13r^3-2r_{Q0}^3)j'  \cr
  - & 156r^6H_0^2j + 3r^4H_0 f (13r^3 - 2r_{Q0}^3) h' - 60n_0r^4H_0^2c' \cr
  + & \left[ 2r^\frac52H_0^\frac32(13r^3-2r_{Q0}^3) + \frac{8n_0^2 H_0^\frac12 (11r^3 - 4r_{Q0}^3) }{(5r_H^3 + 4r_{Q0}^3) r^\frac12} \right]
  \p\beta = 0,
\end{align}
and the EOM of $B$ (\ref{EOM: B}) gives
\begin{align}\label{Diff Eq: B}
  2r^3k' + 6r^2k - 2r^3fj' -12r^2j + 3r^3fh' + \left[ 2r^\frac32H_0^\frac12 - \frac{ 4n_0^2 r^{-\frac32}H_0^{-\frac12} } {5r_H^3+4r_{Q0}^3} \right]
  \p\beta = 0.
\end{align}
These 3 equations for scalar perturbations are not independent, as one can check that
\begin{align}
  5\cdot\text{EOM of}~\phi + 8r^4H_0^2\cdot\text{EOM of}~B - \text{EOM of}~A=0.
\end{align}

The (0) component of Maxwell equation gives
\begin{align}\label{Diff Eq: Maxwell(0)}
  \p_r(r^4H_0^2c') - \frac92n_0h' + 3n_0j' = 0.
\end{align}
This equation can help us to eliminate $c'$ from the other equations.

The $(ii)$ component of Einstein equation
\begin{align}
 \sum_i (E_{ii} - T_{ii}) = 0
\end{align}
gives
\begin{align}
  & 6r^8H_0^2k'' + 3r^4H_0(13r^3+16r_{Q0}^3)k' + 9(5r^6+10r_{Q0}^3r^3+8r_{Q0}^6)k   \cr
  +& 12r^4H_0^2(r^4fh')' - 6r^4H_0^2(5r^3-2r_H^3)j' - 90r^6H_0^2j + 18n_0r^4H_0^2c' \cr
  +& \bigg[ 4r^\frac32H_0^\frac32(5r^3+2r_{Q0}^3) + \frac{12n_0^2r^\frac52H_0^\frac32}{5r_H^3+4r_{Q0}^3} \bigg]\p\beta = 0.
\end{align}
We record this equation here just for the completeness of the paper and the convenience of the readers, we will not use it when solving the perturbations.
But it can be used to check the solutions for the perturbations.

The strategy for solving the scalar perturbations is in the order of $h$, $j$, $c$ and $k$. 4 times the second scalar constraint
(\ref{Diff Eq: 2nd Einstein constraint}) plus 3 times the EOM of $\phi$ (\ref{Diff Eq: phi}) minus the EOM of $A$ (\ref{Diff Eq: A}) will give us
\begin{align}\label{h' and j'}
  &4r^4fH_0(5r^3+2r_{Q0}^3)j' - 6r^4H_0\left[ (5r^3+2r_{Q0}^3)f-2(5r^3-2r_H^3)H_0 \right]h' \cr
  &+ 4r^\frac52H_0^\frac32(5r^3+2r_{Q0}^3)\p\beta = 0.
\end{align}
Using this to eliminate $j'$ in the $(rr)$ component of Einstein equation one gets the equation for $h$:
\begin{align}
  \p_r(r^4f\p_r h) = -\frac{5r^3+2r_{Q0}^3}{3r^\frac32H_0^\frac12} \p\beta.
\end{align}
We can see that the l.h.s. is the same as the differential equation for $F(r)$ and the r.h.s., i.e. the source part is $1/3$ of that for differential equation of $F$.
So we can get immediately $h=F/3$. This relation also holds in \cite{Wu1508} for the case of compactified D4-brane. But from the experience of solving the second order perturbations in \cite{Wu1604}, we know that we should solve $h$ and $j$ to the order of $1/r^6$ in order to get the correct term of order $1/r^3$ for $k$, so we record here the solution of $h$ to the order of $1/r^6$ for the readers' convenience. We have $h$ as
\begin{align}
  h=\left( \frac4{3r^\frac12}-\frac{2r_H\sqrt{r_H^3+r_{Q0}^3}}{9r^3}-\frac{2(2r_H^3+r_{Q0}^3)}{21r^\frac72}
  -\frac{r_H^4\sqrt{r_H^3+r_{Q0}^3}}{9r^6} \right)\p\beta.
\end{align}
Then $j$ is easy to get from eq. (\ref{h' and j'}):
\begin{align}
  j = \left( \frac{r_H\sqrt{r_H^3+r_{Q0}^3}}{3r^3}-\frac{12(r_H^3+r_{Q0}^3)}{35r^\frac72}
  +\frac{r_H\sqrt{r_H^3+r_{Q0}^3}(11r_H^3+6r_{Q0}^3)}{30r^6} \right)\p\beta.
\end{align}
Next, we put the results into the (0) component of Maxwell equation (\ref{Diff Eq: Maxwell(0)}) and get $c$
\begin{align}
  c \sim \mathcal O\left( \frac1{r^\frac72} \right)\p_0r_{Q0}.
\end{align}
So $c$ is trivial and will not contribute to the conserved current. Substitute $h$ and $j$ into the differential equation of $B$ one has
\begin{align}
  r^3k = \left( C_k + \frac{20(3r_H^6+4r_H^3r_{Q0}^3+r_{Q0}^6)}{7(5r_H^3+4r_{Q0}^3)r^\frac12}
  -\frac{r_H^4\sqrt{r_H^3+r_{Q0}^3}}{3r^3} \right)\p\beta,
\end{align}
where $C_k$ is the integral constant and can be fixed by the requirement that the boundary stress tensor is in the Landau frame: $C_k=-\frac2{15}$. We
would like to stress that though we record the results for the scalar perturbations to the order of $1/r^6$, only the term of order $1/r^3$ will contribute to the energy momentum tensor of the boundary relativistic fluid.

\section{The stress tensor and conserved vector current of the boundary fluid}\label{sec: result}

In this section we will derive the boundary stress tensor and the conserved vector current for the relativistic fluid on the boundary so that we can read all the thermodynamic and the hydrodynamical quantities.

The total action of our system is eq. (\ref{5D action: total}). The details of deriving the boundary stress tensor can be found in \cite{Wu1508}. One should
notice that from (\ref{5d reduced metrc}) or (\ref{5D boosted background fields}) we can see that both $\phi$ and $A$ depend on $H_0$ and hence $x$, when $r_{Q0}$ is promoted to be $x^\mu$ dependent. Since the boundary is at some $r={\rm const}$, in the case without D0 charge, the scalars are all constant on the boundary. But now with D0-brane present, $\phi$ and $A$ will also vary on the boundary. This means the boundary hyperplane on which the relativistic fluid resides is not ``iso-D0-charged"---the relative number density of D0-brane will fluctuate on the boundary. This is quite different from the charged $AdS_5$ black hole where the boundary stress tensor is still the same as in the case without charge, which can been seen from the eq. (4.35) of ref. \cite{Banerjee0809}. Thus the boundary stress tensor in this D0-D4 plasma should be modified to
\begin{align}
  T_{\mu\nu} = \frac1{2\kappa_5^2}\lim_{r\to\infty}r^\frac53 \cdot 2\left( K_{\mu\nu}-h_{\mu\nu}K
  -\frac52r^{-\frac13}H_0^{-\frac16}\left( 1-\frac{r_{Q0}^2\dta r_{Q0}}{2H_0r^3} \right)h_{\mu\nu} \right).
\end{align}
Put the first order expanded metric (\ref{5D expanded metric}) together with the solutions of its perturbations (\ref{1st order pert ansatz}) in, we get
\begin{align}\label{boundary stress tensor}
  T_{\mu\nu} = \frac1{2\kappa_5^2} \left[ \left(\frac52r_H^3+3r_{Q0}^3\right) u_\mu u_\nu + \frac12r_H^3P_{\mu\nu} - r_H\sqrt{r_H^3+r_{Q0}^3}
   \left( 2\sigma_{\mu\nu} + \frac4{15}P_{\mu\nu}\p_\rho u^\rho \right) \right],
\end{align}
from which we can read the energy density, pressure, shear and bulk viscosity as
\begin{align}\label{stress tensor quantities}
  & \varepsilon = \frac1{2\kappa_5^2}\left( \frac52r_H^3 + 3r_{Q0}^3 \right),~~~~p = \frac1{2\kappa_5^2}\frac12r_H^3, \cr
  &\eta= \frac1{2\kappa_5^2}r_H\sqrt{r_H^3+r_{Q0}^3},~~~~~~\zeta= \frac1{2\kappa_5^2}\frac4{15}r_H\sqrt{r_H^3+r_{Q0}^3}~.
\end{align}
From (\ref{stress tensor quantities}) one can see that the bulk to shear viscosity ratio is not changed: $\zeta/\eta=4/15$, as compared with ref. \cite{Wu1508}.

To calculate the entropy, we use
\begin{align}
  S = \frac{A_H}{4G_5} = \frac1{2\kappa_5^2}4\pi A_H,
\end{align}
where $A_H$ is the area of the horizon. The entropy can then be calculated as
\begin{align}
  S = \frac1{2\kappa_5^2}4\pi\int d^3x\sqrt{g_{\vec x}}\bigg|_{r_H} = \frac1{2\kappa_5^2}4\pi\int d^3x r^\frac52H_0^\frac12\bigg|_{r_H} = \frac1{2\kappa_5^2}4 \pi V_3 r_H^\frac52H_0^\frac12(r_H).
\end{align}
Thus the entropy density is
\begin{align}\label{entropy density}
  s = \frac{S}{V_3} = \frac1{2\kappa_5^2} 4\pi r_H^\frac52H_0^\frac12(r_H).
\end{align}
From the Hawking temperature (\ref{Hawking temperature}) and the above expression, we can see that if the temperature goes to zero $T\to0$, then we should have $r_H\to0$, which makes $s\to0$. This is quite different from the Reissner-Nordstrom case where the entropy does not go to zero when $T\to0$ at the extremal condition. The shear and bulk viscosity to entropy ratios are also not changed comparing with ref. \cite{Wu1508}:
\begin{align}
  \frac\eta s=\frac1{4\pi},~~~~\frac\zeta s=\frac1{15\pi}.
\end{align}

The chemical potential is defined and calculated as
\begin{align}\label{axial chemical potential}
  \mu = A_t(\infty) - A_t(r_H) = \frac{n_0}{r_H^3 + r_{Q0}^3}.
\end{align}
It is smaller than 1 since from eq. (\ref{axial chemical potential}) one has
\begin{align}
  \mu^2 = \frac{n_0^2}{(r_H^3+r_{Q0}^3)^2} = \frac{r_{Q0}^3}{r_H^3+r_{Q0}^3} < 1,
\end{align}
given that both $r_H$ and $r_{Q0}$ are larger than 0. This will be useful when we analyze the stability of the boundary fluid. The conserved vector current is defined as
\begin{align}
  J_\mu = \frac1{2\kappa_5^2}\lim_{r\to\infty} 3r^3A_\mu
\end{align}
with the result is
\begin{align}
  J_\mu = \frac1{2\kappa_5^2}\left(Q_0u_\mu - \frac{r_H^4}{(r_H^3+r_{Q0}^3)^\frac12} T P_\mu^\nu \p_\nu\left( \frac{\mu}T \right) + \frac{4\kappa_{CS}r_{Q0}^3(3r_H^3+r_{Q0}^3)}{3(r_H^3+r_{Q0}^3)^2}l_\mu \right).
\end{align}
Compared with the following definition for the current of relativistic fluid
\begin{align}
  J_\mu = \rho u_\mu - \sigma TP_\mu^\nu\p_\nu\left( \frac{\mu}T \right) + \sigma_V l_\mu,
\end{align}
we then have
\begin{align}\label{axial current quantities}
  \rho = \frac1{2\kappa_5^2}3n_0,~~~\sigma = \frac1{2\kappa_5^2} \frac{r_H^4}{(r_H^3 +r_{Q0}^3)^\frac12},~~~ \sigma_V = \frac1{2\kappa_5^2}\frac{4\kappa_{CS}r_{Q0}^3(3r_H^3+r_{Q0}^3)} {3(r_H^3+ r_{Q0}^3)^2}.
\end{align}
Here the $\sigma_V$ is the CVE coefficient for the D0-D4 plasma. From eqs. (\ref{Hawking temperature}, \ref{stress tensor quantities}, \ref{entropy density}, \ref{axial chemical potential}, \ref{axial current quantities}) one can see that the Smarr relation is satisfied
\begin{align}
  \varepsilon + p = sT + \mu\rho.
\end{align}

Since the temperature and chemical potential both depend on $r_H$ and $r_{Q0}$, thus we can solve $r_H$ and $r_{Q0}$ in terms of $T$ and $\mu$ to reexpress the results. They are listed in table \ref{tab: results in T and mu}.
\begin{table}
\centering
\scalebox{1.25}{\begin{tabular}{|c|c|}
  \hline
  $\varepsilon$  & $\frac1{2\kappa_5^2}\left(\frac{4\pi}3\right)^6 \frac{(5+\mu^2)T^6}{2(1-\mu^2)^4}$ \\
  \hline
  $p$  & $\frac1{2\kappa_5^2}\left( \frac{4\pi}3 \right)^6 \frac{T^6}{2(1-\mu^2)^3}$ \\
  \hline
  $\eta$ & $\frac1{2\kappa_5^2}\left( \frac{4\pi}3 \right)^5 \frac{T^5}{(1-\mu^2)^3}$ \\
  \hline
  $\zeta$ & $\frac1{2\kappa_5^2}\left( \frac{4\pi}3 \right)^5 \frac{4T^5}{15(1-\mu^2)^3}$ \\
  \hline
  $\rho$ & $\frac1{2\kappa_5^2}\left( \frac{4\pi}3 \right)^6 \frac{3T^6\mu}{(1-\mu^2)^4}$ \\
  \hline
  $\sigma$ & $\frac1{2\kappa_5^2}\left( \frac{4\pi}3 \right)^5 \frac{T^5}{(1-\mu^2)^2}$ \\
  \hline
  $\sigma_V$ & {\footnotesize$\frac1{2\kappa_5^2}4\kappa_{CS} \mu^2 \left( 1-\frac23\mu^2 \right)$} \\
  \hline
\end{tabular}
}
\caption{\label{tab: results in T and mu} Reexpress the thermal and hydrodynamical quantities in terms of $T$ and $\mu$.}
\end{table}

From the thermal quantities in table \ref{tab: results in T and mu}, we can derive some other thermal properties. The D0-brane charge number susceptibility is defined \cite{KimYoungman1001} and calculated as
\begin{align}
  \chi = \left.\frac{\p\rho}{\p\mu}\right|_{\mu=0} = \frac1{2\kappa_5^2}\left( \frac{4\pi}3 \right)^63T^6.
\end{align}
It is interesting to compare this result with that of ref. \cite{Erdmenger0809} where $\chi\sim T^2$. There is another very interesting relation between the CVE coefficient and the thermal quantities. From the results of eqs.  (\ref{stress tensor quantities},\ref{axial chemical potential},\ref{axial current quantities}), one can easily see that $\mu=\rho/(\varepsilon+p)$. So we can reexpress the CVE coefficient $\sigma_V$ in table \ref{tab: results in T and mu} as
\begin{align}\label{CVE relate with thermal quantity}
  \sigma_V = \frac1{2\kappa_5^2}4\kappa_{CS} \mu^2 \left( 1-\frac23\frac{\mu\rho}{\varepsilon+p} \right).
\end{align}
From the above result we may draw two important conclusions. The first one is that the above result justifies the CS term that we add manually in
(\ref{5D action: bulk}). The CVE term $\sigma_V l_\mu$ in the conserved current of the boundary fluid comes from the CS term added ``by hand" in the 5D bulk action (\ref{5D action: bulk}). One may wonder whether it is reasonable to do such modifications on the 5D system dimensionally reduced from the background of compactified D4-brane with smeared D0-brane charge. So by calculating the concrete result for the CVE coefficient we can see that the CVE term resulting from our modification is actually allowed or admitted to be present by the thermodynamics of the 5D system. The second one is that the anomalous transport coefficients for strongly coupled relativistic fluid seems irrelevant with the conformality of the fluid. This can be seen by comparing our result (\ref{CVE relate with thermal quantity}) (which is a result for a nonconformal relativistic fluid) with the results in \cite{Son0906,Amado1102} (where the boundary fluid is conformal). We can see that if we do not count in the temperature square part of the CVE coefficient (like the case in \cite{Landsteiner1107}), the anomalous transport coefficients is indeed irrelevant with the conformality of the fluid.


At the end of this paper, we want to prove that the relativistic fluid on the boundary is stable by checking the stability criterions. According to ref. \cite{DiDato1501}, there are two kinds of stabilities for fluid systems. The first one is the thermal stability, whose stability condition can be given as the heat capacities are positive. The second one is the dynamical stability, of which the stability condition is the simultaneous validness of the following 3 conditions:
\begin{align}
  c_s^2>0,~~~~\Gamma_T > 0,~~~~\Gamma_L > 0,
\end{align}
where $c_s$, $\Gamma_T$ and $\Gamma_L$ are separately the sound speed, the attenuation coefficients for the shear and sound mode dispersion relations:
\begin{align}\label{dispersion relation: convention}
  \omega_T &= -i \Gamma_T \boldsymbol k^2, \cr
  \omega_L &= \pm c_s |\boldsymbol k| -i \Gamma_L \boldsymbol k^2.
\end{align}

Firstly, we check the thermal stability conditions. The enthalpy for this system is
\begin{align}
  \mathcal H = \varepsilon+p = \frac1{2\kappa_5^2}\left(\frac{4\pi}3\right)^6\frac{3T^6}{(1-\mu^2)^4}.
\end{align}
In the calculation, the thermal equilibrium value of the relative density of D0-branes $N_0/(V_4N_4)$ is supposed to be a constant. Since one has
$\rho\sim n_0=Q_0/Q_4\sim N_0/(V_4N_4)$, thus the heat capacity at constant volume and at constant pressure should be calculated at fixed $\rho$:
\begin{align}\label{heat capacities}
  & c_V = \left(\frac{\p\varepsilon}{\p T}\right)_{\rho} = \frac1{2\kappa_5^2} \left(\frac{4\pi}3\right)^6 \frac{3T^5(5-\mu^2)} {(1+7\mu^2)(1-\mu^2)^3}, \cr
  & c_p = \left(\frac{\p\mathcal H}{\p T}\right)_{\rho} = \frac1{2\kappa_5^2} \left(\frac{4\pi}3\right)^6 \frac{18T^5}{(1+7\mu^2)(1-\mu^2)^3}.
\end{align}
Given that $\mu^2<1$, the heat capacities are both positive. The ratio of $c_p$ to $c_V$ is
\begin{align}
  \frac{c_p}{c_V} = \frac6{5-\mu^2}.
\end{align}
This ratio is no longer a constant as in the case of compactified black D4-brane \cite{Naji2016}, it depends on the chemical potential.

Secondly, we will check the dynamical stability condition. The sound speed can be calculated from only the thermal quantities at constant $s/\rho$ \cite{DiDato1501} as
\begin{align}\label{sound speed from thermal relation}
  c_s^2 = \left(\frac{\p p}{\p\varepsilon}\right)_{s/\rho} = \left|\frac{\p(p,s/\rho)}{\p(T,\mu)}\right| \bigg/ \left| \frac{\p(\varepsilon,s/\rho)}{\p(T,\mu)} \right| = \frac{1-\mu^2}{5-\mu^2} > 0.
\end{align}
Note that the condition that $s/\rho$ is kept fixed is very important for getting the correct result of the sound speed. The dispersion relation is calculated by working in the linear regime of the fluid \cite{Bhattacharyya0712,Wu1508,Wu1604}. In the circumstances now, the linear regime is achieved by expanding $r_H$, $r_{Q0}$ and $u_\mu$ as
\begin{align}
  r_H(x) = r_H + \dta r_H e^{ikx},~~~~ r_{Q0}(x) = r_{Q0}+\dta r_{Q0}e^{ikx},~~~~u_\mu(x) = (-1,\dta\beta_ie^{ikx}).
\end{align}
Then put the above into the EOM of the fluid\footnote{Since we do not turn on the background field like in \cite{Son0906,Megias1304}, thus the EOMs for stress tensor and charge current are still take their conserved form.}
\begin{align}
  \p^\mu J_\mu &= 0, \label{fluid EOM: current}\\
  \p^\mu T_{\mu\nu} &= 0. \label{fluid EOM: stress tensor}
\end{align}
One can solve out $\dta r_{Q0}$ from eq. (\ref{fluid EOM: current}) in terms of $\dta r_H$ and $\boldsymbol k\cdot\boldsymbol{\dta\beta}$ and then
substitute it into eq. (\ref{fluid EOM: stress tensor}). The condition that the determinant of the coefficient matrix for the vector $(\dta r_H, \dta\beta_i)^T$ is zero gives a quintic algebraic equation for $\omega$ which can be factorized\footnote{The ``T" in $(\dta r_H, \dta\beta_i)^T$ stands for transpose.}. We can solve out the dispersion relation from this quintic equation. The result is
\begin{align}
  &\omega_T = -i \frac{r_H}{3(r_H^3+r_{Q0}^3)^\frac12} \boldsymbol k^2, \cr
  &\omega_L = \pm\frac{r_H^\frac32}{(5r_H^3+4r_{Q0}^3)^\frac12}|\boldsymbol k| -i \frac{2r_H(10r_H^3+13r_{Q0}^3)}
  {15(r_H^3+r_{Q0}^3)^\frac12(5r_H^3+4r_{Q0}^3)} \boldsymbol k^2.
\end{align}
Here we omit the higher orders in $|\boldsymbol k|$ for the sound mode. Compared with eq. (\ref{dispersion relation: convention}) the above dispersion
relations give us
\begin{align}\label{sound speed and attenuations (rh,rq)}
  c_s = \frac{r_H^\frac32}{(5r_H^3+4r_{Q0}^3)^\frac12},~~\Gamma_T = \frac{r_H}{3(r_H^3+r_{Q0}^3)^\frac12},~~\Gamma_L = \frac{2r_H(10r_H^3+13r_{Q0}^3)}{15(r_H^3+r_{Q0}^3)^\frac12(5r_H^3+4r_{Q0}^3)}.
\end{align}
All of the above 3 parameters are positive so that this system is stable under the long wavelength fluctuations of the smeared D0 charge and the local temperature associated with the wavy horizon. As one can check
\begin{align}\label{attenuation with transport}
  \Gamma_T = \frac{\eta}{\varepsilon+p},~~~~\Gamma_L = \frac1{2(\varepsilon+p)}\left( \frac43\eta+\zeta+{4c^2\rho^6D\over\mu^2Tc_s^2c_V^2\mathcal H^2} \right)
\end{align}
considering the results of eqs. (\ref{stress tensor quantities}, \ref{axial chemical potential}, \ref{axial current quantities}, \ref{heat capacities}). Here $D$ is the charge diffusion constant and it relates with the conductivity $\sigma$ by \cite{DiDato1501}
\begin{align}
  \left({\rho T \over \mathcal H}\right)^2D = \sigma T,
\end{align}
and $c$ is defined as
\begin{align}
  c = \left({\p\mu\over\p\rho}\right)_T.
\end{align}
From eq. (\ref{attenuation with transport}) we can also restate the dynamical stability condition in terms of transport coefficients as
\begin{align}
  c_s^2>0,~~~~\eta>0,~~~~\zeta>0,~~~~D~(\text{or}~\sigma)>0.
\end{align}
Again, the above 4 conditions should be satisfied at the same time. Compare with the formulation of the attenuation coefficients in compactified D4-branes \cite{Wu1508}, the form of $\Gamma_T$ is not changed. This is because the shear mode of the dispersion relation reflects only the diffusions of the (traceless) tensor part perturbations which is described by $\eta$. While the sound attenuation $\Gamma_L$ will receive a new contribution associated with charge diffusion constant $D$. We can also reformulate eq. (\ref{sound speed and attenuations (rh,rq)}) in terms of $(T,\mu)$ as:
\begin{align}\label{sound speed and attenuations (T,mu)}
  c_s = \sqrt\frac{1-\mu^2}{5-\mu^2},~~~~\Gamma_T = \frac{1-\mu^2}{4\pi T},~~~~\Gamma_L = \frac{(10+3\mu^2)(1-\mu^2)}{10\pi T(5-\mu^2)}.
\end{align}
One can see clearly now that the sound speed from the dispersion relation (i.e. eq. (\ref{sound speed and attenuations (T,mu)})) is the same as it is got through the thermal relations (i.e. eq. (\ref{sound speed from thermal relation})). Note that with $0<\mu^2<1$, we have $0<c_s^2<1/5$. So adding the smeared D0 charge into the D4-brane volume will slow down the sound speed.

\section{Discussions and outlooks}\label{sec: final}

In this paper, we derive out the CVE coefficient in the 5D background which is reduced from the compactified D4-brane with smeared D0-branes via the BDE formalism of fluid/gravity correspondence. The relativistic fluid corresponds to the 5D bulk is nonconformal with a conserved vector current. The vorticity term contributing to the first order non-dissipative part of this vector current is derived from the CS term that is added manually in the 5D bulk action. We derive all the thermal and hydrodynamical quantities up to first order such as the energy density and pressure, the chemical potential and charge density, the shear and bulk viscosities, the conductivity and CVE coefficient, etc. Except that the sound speed and the CVE coefficient rely only on $\mu$, most of the others depend both on $T$ and $\mu$.

We also talk about the stability for the background of the near extremal, compactified black D4-branes with smeared D0-brane charge. This background can be viewed as a combination of the black Dp-brane and the smeared configuration of D0-branes (in terms of ref. \cite{DiDato1501}). According to ref. \cite{DiDato1501}, the former has stable regime in the parameter space while the latter is not stable. The compactified D4-brane with smeared D0-brane charge raise a new kind of background that is interesting to investigate in the frame of ref. \cite{DiDato1501}. Our discussion is only restricted to the near horizon limit for the D4-branes. The result for the stability analysis can be summarized into one sentence: the D4-branes will dominant the main stability property of this background, while the smeared D0 charge can only change the specific value of the thermal/hydrodynamical quantities but not the general tendency. The results that reexpressed in terms of $(T,\mu)$ in table \ref{tab: results in T and mu} is not the completely field theory language because we still have $\kappa_5$ in those expressions. If one formulates these results into complete field theory language like in \cite{Wu1604}, she/he will find that all the results will be proportional to $N_4^2$ with $N_4$ can be chosen freely. Since the stability conditions are just the (simultaneous) positivity of $c_s^2,\Gamma_T,\Gamma_L$ on which the value of $N_4$ does not have effects. So arbitrary choice for $N_4$ will not affect the stability of the fluid as long as $N_4$ is a very large number such that gauge/gravity duality works.

Another interesting difference from ref. \cite{Erdmenger0809} is that there the chemical potential should be kept small i.e. $|\mu/T|\ll1$ so that the fluctuations on the charge horizon $r_-$ will not exceed the event horizon $r_+$. But in our case it seems that there is no additional requirement that should be imposed on $\mu$. It has a bound that $0<\mu^2<1$ by its definition and no additional conditions is required here. This can be accounted by comparing the emblackening factor $f(r)$ for the Reissner-Nordstrom type black hole and the black hole background in this paper. In the Reissner-Nordstrom black hole, we have $f(r)=1 - \frac{r_0^4}{r^4} + \frac{Q^2}{r^6}$, the location of horizon $r_+$ is determined by both $r_0$ and $Q$ at the same time. Actually, $f(r)=0$ will give us two real roots: the charge horizon $r_-$ and the event horizon $r_+$ and we need $r_+>r_-$ in order to avoid the naked singularity. Thus the event horizon is associated with the charge of black hole. While in our case, the emblackening factor is $f(r)=1-\frac{r_H^3}{r^3}$. The horizon is just at $r=r_H$ and does not relate with $r_{Q0}$. This suggests that the 5D black hole that we use in this paper (\ref{5d reduced metrc}) is not of Reissner-Nordstrom type. That's why here we do not need the requirement that $r_-<<r_+$ as in  \cite{Erdmenger0809}.

There is a recent paper \cite{Cai1608} which also calculates the dispersion relation in the same holographic model. Its result for the sound mode attenuation is
\begin{align}
  \Gamma_L^{\text{\cite{Cai1608}}} = \frac1{5\pi T} + \frac{2Q_0}{5\pi T}.
\end{align}
Since this paper does not take into account the vector perturbation, so we should use only the tensor and scalar part contributions in $\Gamma_L$ to compare:
\begin{align}
  \frac1{2(\varepsilon+p)}\left( \frac43\eta+\zeta \right) = \frac{1}{5\pi T} - \frac{\mu^2}{5\pi T},
\end{align}
with $\mu^2=\frac{Q_0^2}{9(r_H^3+r_{Q0}^3)^2}$. The first term is the same, because this is the case without D0 charge. Except the first term, our result is different from that of \cite{Cai1608} even after ignoring the vector part contribution. This can be understandable that \cite{Cai1608} uses the approximation that $Q_0$ is very small and it has expanded in terms of $Q_0$ when solving the differential equations (This can be seen from its section 5). But all the results in our paper are exact, we do not use any approximations.

Considering the achievement that we have made in this paper, there are still some interesting aspects valuable to explore. Firstly, we can add the background gauge field and try to extract $\sigma_E$ and $\sigma_B$. We can also move to the second order for such a construction like ref. \cite{Megias1304}. Secondly, there is another holographic QCD model which is similar like ours, which is called the D-instanton-D3 model. It is the black D3-brane with smeared D(-1)-brane charge \cite{Gwak1203}. This model has a vacuum pseudoscalar field which may have potential usage in mimicking the QCD plasma. Thirdly, ref. \cite{DiDato1501} studies a kind of unstable background consists of only smeared D0-branes with the smeared dimension $p$. As far as we know, there is no work on the second order transport properties of an unstable relativistic fluid at present. Though its significance is not clear now, it is still an interesting trial. Lastly, one may extract the anisotropy of QCD plasma using the solution of smeared Dp-brane in string theory \cite{Lu0409,Lu0503}. A prototype on this direction is ref. \cite{Gahramanov1203}. Though a Dp-brane smeared on one of its transverse directions is not stable \cite{Bostock0405}, it may still have potential use in QCD plasma. For example, a black D2-brane smeared on a transverse spatial direction may be used to describe the anisotropic QCD plasma.

\section*{Acknowledgement}
We thank Danning Li, Yu Lu, Yi Yang, Zhiguang Xiao and Jiaju Zhang for valuable discussions. C. Wu would like to thank the hospitality of the Wigner Research Center for Physics, Hungarian Academy of Sciences since his stay from November 5th 2016. This work is supported by the NSFC under Grant No. 11621131001 and 11275213 (CRC 110 by DFG and NSFC), CAS key project KJCX2-EW-N01, and Youth Innovation Promotion Association of CAS.


\providecommand{\href}[2]{#2}\begingroup\raggedright\endgroup

\end{document}